# How an AI-ready National Data Library would help UK science

Albert Meroño-Peñuela[a,b], Joe Massey[a], Andrew Newman[a], Elena Simperl[a,b]

[a] Open Data Institute, 90 York Way, London, UK
[b] King's College London, 30 Aldwych, London, UK

## Introduction

In this paper, we provide a technical vision for key enabling elements for the architecture of the NDL with a strong focus on building it as an AI-ready data infrastructure through standardised vocabularies, automated analysis tools, and interoperability services. We follow ODI's Multilayer Interoperability Framework (MIF) for data stewardship, covering central socio-technical aspects for the NDL including user-centric approaches to service design and governance.

Data interoperability is the ability of different IT systems to exchange and use data meaningfully, despite differences in formats, standards, quality and underlying technologies. It is crucial for organisations to work together seamlessly. It is not easy to achieve: aside from multiple data formats and standards, differences in data models and quality, and incumbent data silos, there are privacy and security concerns, as well as organisational, cultural, and legal barriers. UK open data is of high quality, reusable by researchers, and of varied domains (economy, crime, defence, education, health, culture, towns and cities, etc.). The value of this data for research is out of question; however, datasets are still difficult to find, access, and combine seamlessly. This hampers the discovery of fundamental knowledge and affects the UK's position as a science and AI superpower, as outlined by the government's AI Action Plan.[1]

To address these issues, the government is pursuing the ideas of a **National Data Library (NDL),** which will *"bring together existing research programmes and help deliver data-driven public services, whilst maintaining strong safeguards and ensuring all of the public benefit."* [2] Although the specifics of NDL still need to be agreed, the goal of the NDL is to allow for data across government to be used seamlessly for scientific, social, and economic benefits. This is a daunting task, but many building blocks to address it already exist and we believe efforts should focus on integrating them rather than on (re)building them from scratch. For example, federation is an established method for sharing privacy-sensitive datasets in a trustworthy way across organisations for which e.g. ActivityPods (which combine the W3C ActivityPub standard and the Solid specification) can be used. Here we recommend such building blocks, based on our previous work on "user-centric design, curation of high-value datasets, open standards, federation, and interoperability", and ways to combine them for the NDL.

---

[1] https://www.gov.uk/government/publications/national-ai-strategy-ai-action-plan/national-ai-strategy-ai-action-plan
[2] https://labour.org.uk/change/kickstart-economic-growth/

However, AI is changing the ways people work with data. This has an impact in the ways the think about the NDL. Foundational models (e.g. stability.ai's [Stable Diffusion](#) for images; OpenAI's [GPT-4](#) for text; Meta's [MusicGen](#) for music) are a step change across sectors and applications, including researchers and scientists which are key prospective users of the NDL. For researchers, traditional data analysis and data science tools methods have become part of the repertoire of AI—for example, in tasks like [discovering, engineering](#), and [representing](#) knowledge. This means AI plays now an active role in data preparation, e.g. in data cleaning and [augmentation](#). But beyond this, we need to imagine the possibilities of **AI-based scientific discovery tools that use datasets and models hosted in an AI-ready NDL**. Recent studies show that the [use of AI in scientific research is widespread and racing ahead across science](#) since 2015 and is increasingly used to perform research tasks.

Modern AI is trained with datasets, and it has been shown that [AI models are only as good as the datasets they are trained with](#). This means that the NDL must **not just be ordinarily data-ready, but to be AI data-ready.** AI readiness can be understood as a combination of data stewardship good data practices (i.e. as the [ODI defines](#), "collecting, maintaining and sharing data" and making important decisions about "who has access to it, for what purpose, and to whose benefit") and AI. Good data stewardship practices are summarised in the FAIR Guiding Principles for scientific data management and stewardship.[3] For data to be [FAIR](#), it has to be Findable, Accessible, Interoperable and Reusable. Findability is about providing rich and indexed metadata with persistent identifiers. Accessibility requires datasets to be served using standard communication protocols. Reusability asks data publishers to attach appropriate licensing and provenance information to their datasets. At the ODI we extend the FAIR notions of Interoperability with the [Multilayer Interoperability Framework](#) (MIF). The MIF provides four dimensions to analyse and co-design systems that balance provision of value, ensure interoperability, and enable protection against harm:

- the **technical stack,** which extends the FAIR idea of Interoperability (i.e. data must be able to integrate with other data using shared knowledge representation languages and vocabularies) by providing an ecosystem of services that can be adapted to various use-cases, and that facilitates the collection of data, making it AI ready, its access, and its sharing;
- the **governance model**, which answers questions such as who makes decisions and how are the public involved with a user-centric approach;
- **legal forms**, which sustain the design of the NDL in the legal system of the UK; and the
- **commercial model**, which describes the high-level economics of building and maintaining an AI-ready NDL.

Although all these principles are essential for data reuse, they are not enough for ensuring their use for training and building safe and trusted AI systems, because these require datasets to be of specific sizes, meet specific distributions, be free from errors or incomplete data, etc. Therefore, AI-readiness focuses on the documentation of data processes towards its use in AI, such as data preparation (labelling, outliers), data quality (bias, completeness, consistency, integrity), data documentation (metadata, previous uses, intended uses, feedback), and data access (formats, delivery, privacy, security). If the NDL is meant to become a central UK hub for cross-governmental, interoperable data that facilitates scientific discovery using safe and trustworthy AI, then it needs to be an AI-ready NDL.

---

[3] https://www.nature.com/articles/sdata201618

In this paper, we focus on the technical stack and the governance model of the MIF. The rest of the paper is organised according to these two dimensions: we look first at the ecosystem of technical elements for the NDL (data documentation, data assurance, benchmarks and dashboards, data sources, and AI-enabled interoperability), and then propose a user-centric governance model.

## Technical stack: an ecosystem of AI-enabling elements for the NDL

Figure 1 shows a conceptual blueprint for key enabling elements of the architecture of NDL that are essential for enabling its AI-readiness. We conceptualise this as two layers, with AI-readiness as an overlay over the NDL. We focus here on the elements of the AI readiness layer because they drive the requirements for the whole NDL, suggesting specific scenarios and implementations for UK researchers. These elements are:

- **Metadata & dataset documentation**, which describe datasets broadly according to various relevant metadata formats and standards, and enact a solid ground for achieving data interoperability;
- **Analysis**, which contains services and tools that examine datasets at a deep level inferring statistical distributions, outliers, biases, dataset uses, links, etc and lifting this information to enrich metadata & documentation;
- **Benchmarks and dashboards**, which assess the AI-readiness of underlying datasets through the execution of metrics, KPIs, visualisations, etc. and provide a global decision-level view of how much AI-ready datasets in the NDL are;
- **Open standards**, which are orthogonal to all AI-readiness services.

At a more basic level, the infrastructure of the NDL needs to meet the principles and capabilities of the National AI Strategy. In this paper we focus on the capabilities: understanding the long-term needs of the AI ecosystem, focusing on the technical aspects of AI and its infrastructure requirements; and effective AI governance.

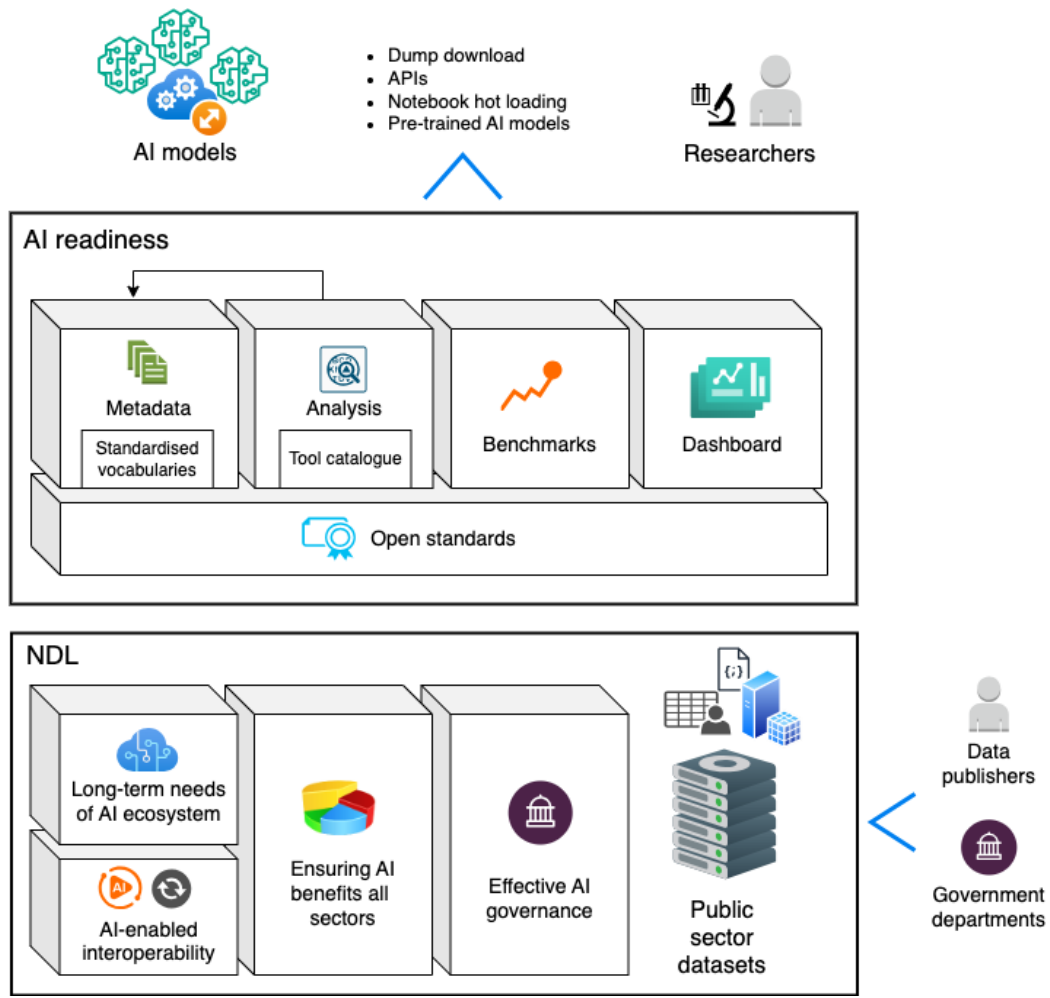

*Figure 1. Conceptual architecture with key elements of the ecosystem of an AI-ready NLD.*

## Documenting data, data properties, and data processes

The first element of AI-readiness (Fig. 1) is metadata. Our vision is based on describing AI datasets with machine-readable, structured metadata as provided by standardised vocabularies, because that facilitates finding, measuring, and integrating datasets.[4] We focus here on vocabularies for **describing datasets**, enriching them with discoverable **links**, making their **privacy** constraints and other **access control** and digital rights explicit, enabling a data-centric description of **AI risks**, allowing **user feedback** to be expressed in structured ways, and representing **machine learning processes** associated with data.

Metadata and data standards have a long history of successes in addressing data interoperability across distributed datasets, mostly centred around ontologies and vocabularies that have been produced over the last 20+ years by the Semantic Web community and the W3C. Many of these standards are directly relevant and should be adopted by the NDL at the dataset metadata level. For example, the Data Catalogue Vocabulary (DCAT) is a "vocabulary designed to facilitate interoperability between data catalogues published on the Web"; this allows publishers of datasets in NDL " to describe datasets and data services (...) using a standard model and vocabulary that facilitates the consumption and aggregation of metadata from multiple catalogues. This can increase the discoverability of datasets and data services" and enables federated search using the same query mechanisms. DCAT is used in UK government data portals (e.g. data.gov.uk) and is endorsed by the Data Standards Authority. An example in RDF syntax:

```
<dataset>            dct:title   "Spend over £500" ;
                     dct:description   "Spend transactions published monthly
according to the Treasury transparency guidelines." ;
                     dct:identifier    <http://dx.doi.org/10.7927/H4PZ56R2> ;
                     dct:license<http://www.nationalarchives.gov.uk/doc/open-
government-licence/version/3/> ;
                     dct:publisher     [
                         rdf:type     foaf:Organization ;
                         foaf:name    "Geological Society" ;
                         foaf:mbox    "info@gs.org" .
                     ] .
```

---

[4] Hogan, A., Blomqvist, E., Cochez, M., d'Amato, C., Melo, G.D., Gutierrez, C., Kirrane, S., Gayo, J.E.L., Navigli, R., Neumaier, S. and Ngomo, A.C.N., 2021. Knowledge graphs. *ACM Computing Surveys (Csur)*, *54*(4), pp.1-37.

Another example in an alternative data.json file:

```
[
  {"title": "Live traffic information from the Highways Agency",
  "license": "No license provided",
  "publisher": {"name": "Highways Agency", "mbox": "test@test.com"},
  "distribution": [
    {
      "downloadURL": "https://s3-eu-west-
1.amazonaws.com/lmtesting2810/HATRIS_15MinuteMIDAS_YYYY-MM-DD_0.csv.zip",
      "title": "Hatris 15 Min Midas 0",
      "format": "application/zip"
    }]
  },
  {"title": "Roadworks locations",
    ...
  }
]
```

Summarising all these standards stand the principles of Linked Open Data: by providing open data on the web, in machine readable formats, using open standards, and a knowledge representation language like the Resource Description Framework (RDF), datasets can be represented in an interoperable way, allowing their interconnection via inter-dataset links and allowing AI users to discover data related to theirs that can expand their applications and research. The ODI provides learning resources on publishing open linked data that will be crucial for the design of an AI-ready NDL that ensures interoperability and provides data discovery mechanisms to its users.

The use of these vocabularies in government, geospatial and healthcare data is very widespread,[5] thus giving the NDL a good opportunity for bootstrapping these domains (see Data Sources). Privacy is a central element of NDL, and especially for health data a wide range of standards could be adopted by NDL to enhance access control and monitoring; this is especially relevant for high-risk AI applications.

Not all data in the NDL will be open, and for datasets with protected information it will be necessary to use structured vocabularies identifying data controllers, describing collection processes, purpose, etc. as demanded by regulatory frameworks (UK's Data Protection Act). A good candidate is the Data Privacy Vocabulary (DPV) which "enables expressing machine-readable metadata about the use and processing of personal data based on legislative requirements"[6] By supporting it, the NDL can provide researchers with more systematic ways of finding and combining datasets; and means for at least partly automating compliance assessment processes that typically entail a lot of manual work. The code below shows an example of a data controller, "Acme", that collects user emails for the provision of services:

```
ex:Acme rdf:type dpv:DataController .
ex:AcmeMarketing rdf:type dpv:Process ;
    dpv:hasDataController ex:Acme ;
    dpv:hasPersonalData pd:EmailAddress ;
    dpv:hasProcessing dpv:Collect, dpv:Use ;
    dpv:hasPurpose dpv:ServiceProvision .
```

The NDL also needs to consider the needs and requirements of data publishers and stewards, who will decide and control under which licenses they publish their data and what rights they wish to grant or retain. Languages such as the Open Digital Rights Language (ODRL) provide vocabularies for the "use of digital content in publishing, distribution, and consumption of digital media", supporting the expression of rights for commercial transactions and public/open content. In the following example in RDF format, a policy specifies that an asset has permission to be reproduced and prohibition to be modified:

---

[5] https://lod-cloud.net/
[6] https://eur-lex.europa.eu/eli/reg/2016/679/oj

```
<http://example.com/policy:0099>
                a odrl:Set;
                odrl:permission [
                    a odrl:Permission ;
                    odrl:target <http://example.com/asset:9898> ;
                    odrl:action odrl:reproduce
                ] ;
                odrl:prohibition [
                    a odrl:Prohibition ;
                    odrl:target <http://example.com/asset:9898> ;
                    odrl:action odrl:modify
                ] .
```

The strong dependence between AI behaviour and the data that is used for its training raises the need to have a more data-centric risk assessment, supported by machine-readable vocabularies. An AI-ready NDL needs to consider the documentation of AI risks especially in the context of AI data. For example, the AI Risk Ontology (AIRO) allows representing AI-related risks, determining which of them are high-risk, and e.g. automatically producing relevant documentation for legal compliance. AIRO is not a perfect fit here as it is not data-centric and focuses on legal requirements; but it is a starting point, as it can enable semi-automated compliance, potentially saving public resources. Modern machine-readable vocabularies for data-centric AI risk assessment are, however, still needed.

For the NDL to be sustainable it needs to be maintained, and long-term maintenance of a successful online platform needs feedback from users.[7] This feedback comes in all forms and shapes: for example, Wikipedia and Wikidata employ free text discussion pages for decisions, including sustainability. However, unstructured user feedback is difficult to search and measure (e.g. to study what datasets are more frequently accessed or which contain more urgent issues to fix), and structured machine-readable metadata is much more appropriate. A good vocabulary for this is the Dataset Use Vocabulary (DUO), providing mechanisms for platform users to "describe consumer experiences, citations, and feedback about the dataset from the human perspective", and enabling a feedback loop for user-centric design that is a core element of our vision of AI-readiness for the NDL (see below: Governance model: a user-centric approach). Interoperable feedback annotations in DUO can be used by NDL maintainers to understand user behaviour across the platform, frequent workflows and pathways, recurrent queries, new requirements, etc. If they are automatically saved and appropriately anonymised, query logs can be used to better understand user needs.[8]

Finally, various vocabularies have been proposed to describe AI data and its processing. Some of them do so focusing on human-readability, like Data cards (in a semi-structured format) and similarly Datasheets for Datasets (unstructured), which facilitate the documentation of a dataset's motivation, composition, collection process, recommended uses, data subjects, versioning information, etc. Other approaches focus on more structured, machine-readable formats:

1. Croissant, maintained by the MLCommons community, is the de facto standard adopted by Google Dataset Search, Hugging Face, and Kaggle. It is a vocabulary based on schema.org that includes "key attributes and properties of datasets, as well as information required to load these datasets in ML tools" and thus enabling "interoperability between ML frameworks" and making ML datasets easier to find, understand, and use. It does so through four layers: the *metadata layer* which provides the dataset name, provenance and licensing; the *resources layer* which documents dataset files and contents; the *structure layer* which organises and structures the dataset's contents; and the *semantics layer* which formally describes the meaning of dataset columns and variables for interoperability.
2. MLDCAT-AP describes machine learning models and their associated datasets and processes; proposed by SEMIC Support Centre for a more interoperable Europe administration and OpenML.
3. FAIR4ML is proposed by the Research Data Alliance and extends schema.org for creating machine-readable representations of machine learning models

---

[7] Kraut, R.E., 2012. *Building Successful Online Communities: Evidence-based Social Design*. MIT Press.
[8] Courage, C. and Baxter, K., 2005. Understanding your users: A practical guide to user requirements methods, tools, and techniques. *Gulf Professional Publishing*.

The provision of metadata through these standardised vocabularies in the NDL will ensure its AI-readiness and provide accountability and traceability as required by AI regulatory frameworks (UK's AI regulation white paper, UK's Data Protection Act).[9]

## Assuring data and documenting the results

The basic idea of **analysis** (Fig. 1) is to provide a **tool catalogue** that supports NDL researchers in inferring additional metadata from the actual contents of the datasets dynamically, rather than relying on static, publisher provided values. The performance of AI models depends strongly on data characteristics, and so the more the NDL knows about its datasets, and the more it documents them in an explicit and structured way, the more we can ensure AI models perform as intended.

A good illustrative example for this layer is Croissant RAI, which extends the vocabulary of Croissant by providing additional terms that facilitate the description of ML datasets for responsible AI use and its various use cases. The most relevant of those for this AI-readiness layer are:

- Data life cycle, including motivation, composition, collection process, preprocessing/cleaning/labelling, uses, distribution, and maintenance[10]
- Data labelling concerns information about the labelling process and how data annotations were created, the sample the labels apply to, and the demographics of the people who annotated it (in e.g. crowdsourcing platforms)

---

[9] Hardinges, J., Simperl, E. and Shadbolt, N., 2024. We must fix the lack of transparency around the data used to train foundation models. *Harvard Data Science Review (Special Issue 5)*. https://doi.org/10.1162/99608f92. a50ec6e6.
[10] Timnit Gebru, Jamie Morgenstern, Briana Vecchione, Jennifer Wortman Vaughan, Hanna M. Wallach, Hal Daumé III, Kate Crawford: Datasheets for datasets. Commun. ACM 64(12): 86-92 (2021)

- AI safety and fairness evaluation involves understanding the potential risks and fairness aspects associated with data usage and to prevent unintended and potentially harmful consequences (e.g. known biases of the data)

Many of the properties arising from these use cases are structured, statistical descriptions of the variables and values contained in datasets that can be assessed in real time with tools such as Know your data. Assessing these is critical to address issues of bias. Know your data takes a UI-driven stance, allowing users to "explore datasets, improve data quality and mitigate bias issues" as they examine datasets through:

- Dataset visualisations, e.g. showing pictures and instances of the dataset
- Stats, providing global statistics of the features of the dataset
- Item, with metadata associated to a dataset record
- Relations, showing correlations between different features

Researchers can search, filter, sort and group data and observe statistical properties in their datasets. This opens possibilities on how researchers can flag AI data issues by interacting directly with NDL datasets. For example, a researcher may log in into the NDL, combine various related datasets, refine them (see below: [OBJ], use a Know-your-data equivalent to find an underrepresented group,[OBJ][11] and report the issue using [HYPERLINK "https://research.google/blog/croissant-a-metadata-format-for-ml-ready-datasets/"Croissant.

Extending the notion of being data-ready and data-driven, AI-readiness focuses on the documentation of AI-centric processes that are relevant to the data. A useful way to think about the difference between the two is to consider the Checklist to Examine AI-readiness for Open Environmental Datasets by the Data Readiness cluster of the ESIP (Earth Science Information Partners). Some aspects of the checklist overlap with ML data documentation initiatives mentioned above, but most of them are new and appear when we consider how the statistical characteristics of datasets can be observed, learned, replicated and magnified by AI models that use them for training. Elements of the checklist include:

- **Data Preparation** is the "act of manipulating (or pre-processing) raw data (which may come from disparate data sources) into a form that can readily and accurately be analysed".[12] This includes deciding a policy for missing data (e.g. removing full records, replacing them with the mean of variable, etc.), deciding what to do with outliers (data points that are significantly different from the rest of the data), accounting for how many sources the data was compiled from, the regularisation or normalisation procedure (i.e. making sure all values are comparable and sit within the same range). This is essentially a full provenance record of what was done to the data before publishing it.

---

[11] Abián, D., Meroño-Peñuela, A. and Simperl, E., 2022, October. An analysis of content gaps versus user needs in the wikidata knowledge graph. In *International Semantic Web Conference* (pp. 354-374). Cham: Springer International Publishing.

[12] https://www.iri.com/blog/business-intelligence/a-fresh-look-at-data-preparation/

- **Data Quality** comprises its completeness (share of the total spatial, temporal, demographic, etc. breadth of the data), consistency (how uniform the dataset is compared to similar datasets), bias (a systematic, unfair sampling of the overall data to be represented), timeliness (frequency of data publication compared to the occurrence of the observed phenomenon), provenance (a full history of who/what agents performed what activity on what parts of the data and when) and integrity (whether the data remains unchanged with respect to the original).
- **Data documentation** is covered by the various standards mentioned above (DCAT, Croissant, etc.), but emphasises what previous uses have been made of the data (part of provenance), who are the responsible parties, and what is the purpose and intended uses of the dataset.
- **Data access** includes formats (e.g. CSV, JSON), delivery options (direct file download, API), license and usage rights (who is allowed to use the data, under which constraints or fees) and security and privacy (i.e. protection over data that is limited in some way).

## Supporting capabilities: AI readiness benchmarks and dashboards

By using **Metadata** and **Analysis** (Fig. 1), we know how to have richly documented, AI-ready datasets. However, AI data documentation is not enough for AI-readiness: AI models function as black-box systems with a lack of transparency that is potentially problematic to researchers. For this, we propose here two additional capabilities: **benchmarks**, needed for comparing the performance of different AI models and datasets hosted in the NDL; and **dashboards**, needed for assessing model transparency and explainability.

The idea we propose here is that the NDL does not limit itself to the publication of datasets, but that by **acting as a sandbox it can also host a limited number of AI models pre-trained with NDL data that can be constantly evaluated** according to these metrics, benchmarks and dashboards. In this way, the NDL AI sandbox can tell by means of direct experimentation if the use of its datasets for training AI models may entail harms or undesired outcomes. It can also use them to automatically annotate them with relevant structured vocabularies (see above: [Documenting data, data properties, and data processes](#)).

Benchmarks are necessary to achieve this vision. They are standardised tests used to assess AI system performance under different dimensions; and they are primordial to orient AI, as their design choices encode values and priorities that specify directions for the AI community to understand and improve technology.[13] A good example of AI benchmark is the [Holistic Evaluation of Large Language Models](#) (HELM), which evaluates AI models in different scenarios, tasks, datasets, and metrics relevant to AI readiness (e.g. copyright infringement, disinformation, social bias, toxicity). The NDL wants to invest in them because benchmarks can influence the trajectory of underlying AI models and technology, for example measuring their effectiveness for data **integration**, **interoperability**, or measuring **how much value government datasets can provide to researchers and science**. Existing benchmarks may be costly to implement, and they are generally not independent.

---

[13] Liang, P., Bommasani, R., Lee, T., Tsipras, D., Soylu, D., Yasunaga, M., Zhang, Y., Narayanan, D., Wu, Y., Kumar, A. and Newman, B., 2022. Holistic evaluation of language models. *arXiv preprint arXiv:2211.09110*.

Dashboards are used in governance to provide live visualisations of relevant metrics (e.g. KPIs) and supporting decisions. For the NDL, benchmarks can provide a reliable assessment of the trustworthiness of hosted AI models and datasets. AI models typically function as black-box systems with a lack of transparency that is problematic to users. Similarly to the design of dashboards for transparent conversational AI,[14] we propose that the NDL infers dashboards that accompany AI models and explain their conceptualisations, internal states, and user models in real time. Providing researchers with such internal models of hosted AI systems would improve transparency and explainability, which are fundamental in the scientific process.[15]

## Other key elements

### Data sources, APIs, and models consuming data

As shown in Figure 1, data has various ways of entering (input) and existing (output) the NDL. Raw data sources can be thought of as dataset deposits by data publishers and government departments; however, these datasets may need the most basic data stewardship treatments, for which AI can greatly help (see below: AI-enabled interoperability and harmonisation services). Once AI-ready, we foresee data leaving the NDL in various ways, according to different needs of users: via a dump download (where possible), via APIs (e.g. by using services automating their construction), and via hot loading into data science and analysis platforms (e.g. Jupyter Notebooks) and pre-trained AI warm models (e.g. in Hugging Face, or the NDL AI model sandbox itself).

We believe having an early MVP for the NDL is key. First and foremost, it allows to quickly test its acceptance among researchers, and allow them to provide early feedback; focus on core functionalities; and attract users as a means of creating value. But an early MVP requires making choices on domains. As explained in our recent consultation response on building an AI-ready National Data Library, the success of the NDL depends to a large extent on the "quality, relevance, and accessibility of the data it stewards". We recommend feeding the NDL initially with datasets from these three areas, as they would be a good cross-section of different data types and technology:

1. High-quality public data (e.g. data.gov.uk)
2. Federated Trusted Research Environments (TREs) (e.g. NHS Digital's TRE, INSIGHT Health Data Research Hub)

---

[14] Chen, Y., Wu, A., DePodesta, T., Yeh, C., Li, K., Marin, N.C., Patel, O., Riecke, J., Raval, S., Seow, O. and Wattenberg, M., 2024. Designing a Dashboard for Transparency and Control of Conversational AI. *arXiv preprint arXiv:2406.07882*.

[15] Markus, A.F., Kors, J.A. and Rijnbeek, P.R., 2021. The role of explainability in creating trustworthy artificial intelligence for health care: a comprehensive survey of the terminology, design choices, and evaluation strategies. *Journal of biomedical informatics*, 113, p.103655.

3. Cultural heritage data (National Archives, Museums, catalogues such as [Museum Data Service](), etc.)—rich in people and places and capturing the diversity of the UK

These areas sample the diversity of different types of government data, the varied needs of scientists in finding and accessing them, and the different workflows around curating and publishing them. Additionally, they all use different technologies, are of high-value, are fairly well-structured, and can be reliably deployed in public sector projects without major changes in their governance or underlying technology stack. TREs like [OpenSAFELY]() provide strong guarantees for research data if it cannot be made public.

## AI-enabled interoperability and harmonisation services

As mentioned before, building an AI-ready NDL does not just mean to serve government datasets for researchers in a way that they are prepared to be safely and responsibly used in AI models, but also to ensure **AI is used within the NDL itself to solve data interoperability** issues. This is shown at the bottom-left corner of the NDL platform diagram of Figure 1, encapsulating AI interoperability and harmonisation services that NDL users can use to perform basic data stewardship tasks with the help of AI. We now look in detail into this corner, and unpack these services in Figure 2.

Figure 2 shows a blueprint of how different AI-based services could work together in the NDL to support interoperability and harmonisation processes. The ODI has previously shared views about how AI could help addressing data interoperability on [How to build a National Data Library](), [an AI-ready National Data Library](), and our recent white paper on building a better future with data and AI.[16] The idea here is that despite researchers (in natural science, social science, health sciences, etc.) have different data practices and needs, with **AI assistance their analyses can be done with less technical knowledge on data**. If data interoperability issues can be at least partly solved by an ecosystem of AI services, scientists can then focus more on expanding and integrating their datasets with others and in pursuing deeper, unforeseen research questions.

---

[16] https://theodi.cdn.ngo/media/documents/Building_a_better_future_with_data_and_AI__a_white_paper.pdf

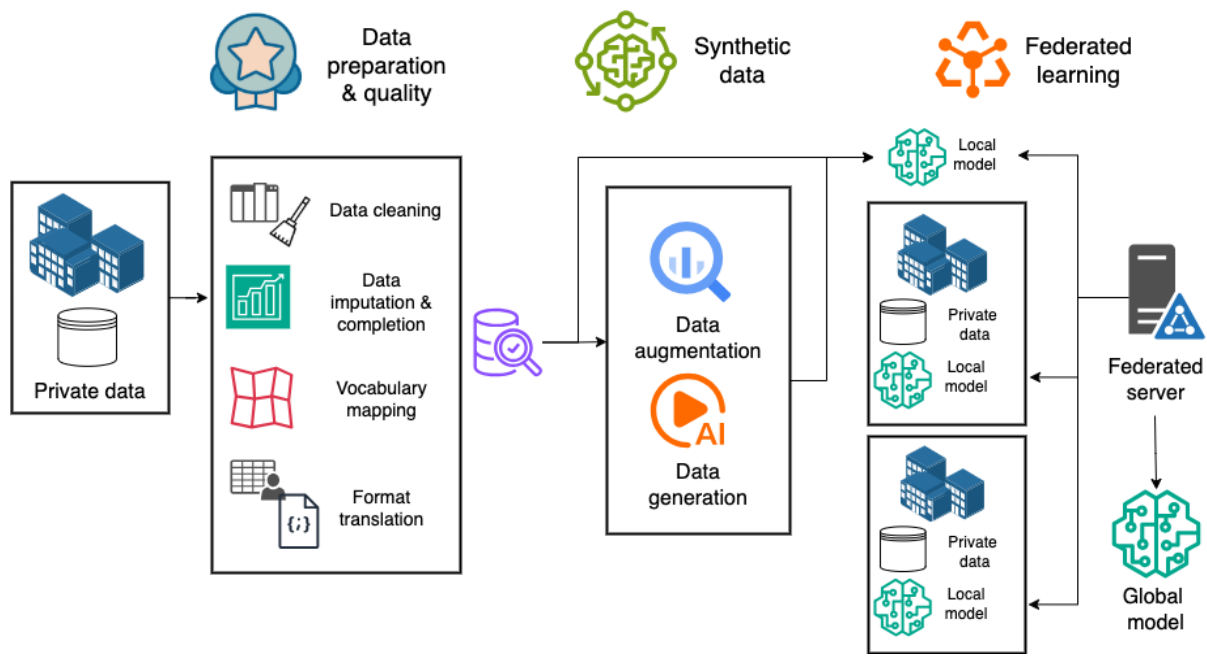

*Figure 2. AI-enabled data interoperability and harmonisation processes through: (1) data preparation and quality; (2) synthetic data; (3) federated learning.*

First, AI can be used as a tool within the NDL to perform basic, routine data stewardship, maintenance, preparation, and quality tasks towards improving its interoperability and integration (Fig. 2.1):

- **Data cleaning** detects "incomplete, incorrect, or inaccurate parts of the data and then replacing, modifying, or deleting the affected data";[17] large language models (LLMs) can be used to identify duplicate or too similar records.
- **Data imputation** replaces missing values with substituted ones, when records are too valuable to be removed when they are incomplete or erroneous. Imputation can technically be done in various ways, from using the variable's mean as a substituted value, statistical regression, etc.
- **Vocabulary mapping** discovers alignments between vocabularies using different terms to represent the same property (e.g. Dublin Core's and Schema.org's properties for "publisher"), even when properties may use synonym words, broader/narrower terms, etc. LLMs can perform ontology and vocabulary alignment semi-automatically.
- **Format translation** transforms information between multiple unstructured (text, images), semi-structured (annotations), and structured (tabular formats, JSON, etc.) formats, for which LLMs are notably good.[18]

---

[17] Wu, S., 2013. A review on coarse warranty data and analysis. *Reliability Engineering & System Safety*, *114*, pp.1-11.
[18] White, J., Fu, Q., Hays, S., Sandborn, M., Olea, C., Gilbert, H., Elnashar, A., Spencer-Smith, J. and Schmidt, D.C., 2023. A prompt pattern catalog to enhance prompt engineering with chatgpt. *arXiv preprint arXiv:2302.11382*.

Second, generative AI like LLMs can perform data augmentation and generate realistic synthetic data (Fig. 2.2).[19] Data augmentation artificially generates new data from existing data, mainly to comply with ML model and AI requirements of being trained with sufficiently large and varied datasets. Real data with such characteristics may be difficult to obtain (due to data silos, economics, legal, etc), and data augmentation can increase both size and variety by making small changes to the original datasets. We foresee the need of expanding current metadata standards (see above: Documenting data, data properties, and data processes) to document these augmentation processes (e.g. "30% of instances of this dataset where created/enriched by Llama").

Third, federated learning allows the training of global AI models (i.e. that go beyond the statistical properties of small, private datasets) on private datasets without compromising their privacy and access policies (Fig 2.3). This is done by a federated server sending the model to different, distributed stakeholders, which never send private data outside their environments. There, they partly train a small, local model and send back the weights to the federated server, which then merges weights received from all federated endpoints; this goes on until convergence is achieved and a global AI model is available. The NDL could facilitate both the federated learning infrastructure and serving secure hosting environments to stakeholders.

Besides the obvious automation they provide over stewardship tasks, these AI services have two additional advantages: (a) they can be used without expert knowledge in statistics that would be usually required without AI; and (b) they typically work well for data modalities that have been traditionally subject less to data curation, such as blocks of text (with LLMs) or images (with e.g. text-to-image models) thanks to the low level, continuous tensors that AI uses in its internal representations.

## Governance model: a user-centric approach

Regarding the governance of data within the NDL, we know that organisations that steward data make important decisions about how data is accessed, used, and shared, including who has access to data, for what purposes and to whose benefit. Governance of data can look different across different organisations and diverse types of data and require different mechanisms to ensure that data is accessed, used, and shared in the right way.

We know that a user-centric approach will be key to co-designing something that delivers to its promise. This includes reviewing key sectors, engaging with experts and a diverse range of stakeholders, deciding on strategic use cases, and ultimately identifying the governance model which best suits the aims of the NDL.

The user-centric approach must cover the full development of the governance model for the NDL. To begin, mapping the ecosystem of actors who may utilise the NDL is key, to ensure a full understanding of prospective users, both publishers and users of data. This full understanding of prospective users means that future designs of the NDL can involve the right blend of stakeholders. For example, when designing what datasets should be stewarded by the NDL, there is a need to work with researchers across disciplines (social sciences, health sciences, etc.) to understand which datasets they need for their research.

---

[19] Whitehouse, C., Choudhury, M. and Aji, A.F., 2023. LLM-powered data augmentation for enhanced cross-lingual performance. arXiv preprint arXiv:2305.14288.

One of the complexities of a NDL is around the stewardship of many different datasets, each of which have their own vulnerabilities, and ethical challenges, meaning that each dataset will require different criteria to access and use the data. Some will be openly available, whereas others should only be accessed by certain people for certain reasons. There are many examples of how organisations can assess who should be able to access data and for what reasons. For example, the [INSIGHT Health Data Hub](#) has a data access board comprised of patients, the public and medical experts, to advise on data access requests. Another example, [OPENSafely](#), has a panel which reviews data use every quarter with the view to updating their [researcher policy](#) with any changes based on any misuse. An example of a collaboratively stewarded AI dataset is the [open AI and algorithm registers of Helsinki and Amsterdam](#), which any individual is able to check and curate.[20]

In the case of the NDL, we believe there should be an opportunity for data subjects to be involved in how data is stewarded. With the complexity mentioned above, there must be a minimum standard adhered to by each and every dataset, and each researcher accessing those datasets. To develop this policy, we recommend following a similar model to the [Camden Data Charter](#), which brought together 20 members of the public to design a charter for ethical data use in the council.

For the NDL, this process should be collaborative, across members of the public, researchers and innovators, to ensure the charter balances ethical, privacy and innovation concerns. The process of deliberation on these principles builds trust between the stakeholders, and goes towards ensuring data is used to the benefit of society. Each data access request must adhere to this charter, which can be updated on a regular basis as data is used by researchers for many different projects. For the case of some sensitive datasets, for example health datasets, additional governance processes may be required, however in many cases, the data publisher may already have a process in place. With an AI-ready NDL we can do all this via the automated stewardship we have described in this paper via standards (see above: [Documenting data, data properties, and data processes](#)) and services (see above: [AI-enabled interoperability and harmonisation services](#)): document it with machine readable metadata (e.g. DCAT), use vocabularies like DUO for usage and feedback from stakeholders and researchers, DPV for sensitive datasets, and Croissant for ensuring data is used for the benefit of all.

---

[20] [https://www.adalovelaceinstitute.org/wp-content/uploads/2021/09/Participatory-data-stewardship_Final-report.pdf](https://www.adalovelaceinstitute.org/wp-content/uploads/2021/09/Participatory-data-stewardship_Final-report.pdf)

## Conclusion

In this paper we share the ODI's vision on a future AI-ready National Data Library. We propose an ecosystem of key AI-enabling elements, focussing on some of its technical components and user-centric governance. We strongly believe that a user-centric governance model will help co-design an NDL governed by citizens for the citizen's good and the empowerment of science. Open standards supporting dataset metadata and documentation, analysis, and benchmarking will be fundamental for the NDL, making it capable of providing live dashboards and AI model sandboxes for continuous evaluation. Combining this with data enrichment and data sharing techniques based on AI in key data stewardship activities such as data preparation, augmentation, and federated learning can make a difference in designing a privacy-aware NDL that is ready for the AI age.